\documentclass{article}

\usepackage{microtype}
\usepackage{graphicx}
\usepackage{subfigure}
\usepackage{booktabs} 
\usepackage{amsmath, amssymb}
\usepackage{hyperref}
\usepackage{paralist}
\usepackage{float}

\newcommand{\R}{\mathbb{R}}
\newcommand{\C}{\mathbb{C}}

\usepackage[accepted]{icml2018}

\icmltitlerunning{Improving DNN-based Music Source Separation using Phase Features}

\begin{document}

\twocolumn[
\icmltitle{Improving DNN-based Music Source Separation using Phase Features}

\icmlsetsymbol{equal}{*}

\begin{icmlauthorlist}
\icmlauthor{Joachim Muth}{epfl}
\icmlauthor{Stefan Uhlich}{str}
\icmlauthor{Nathana\"el Perraudin}{sdsc}
\icmlauthor{Thomas Kemp}{str}
\icmlauthor{Fabien Cardinaux}{str}
\icmlauthor{Yuki Mitsufuji}{tky}
\end{icmlauthorlist}

\icmlaffiliation{epfl}{\'{E}cole Polytechnique F\'{e}d\'{e}rale de Lausanne (EPFL), Lausanne, Switzerland}
\icmlaffiliation{str}{Sony European Technology Center (EuTEC), Stuttgart, Germany}
\icmlaffiliation{tky}{Sony Corporation, Audio Technology Development Department, Tokyo, Japan}
\icmlaffiliation{sdsc}{Swiss Data Science Center, EPFL and ETH Z\"urich, Switzerland}

\icmlcorrespondingauthor{Joachim Muth}{joachim.h.muth@gmail.com}
\icmlcorrespondingauthor{Stefan Uhlich}{stefan.uhlich@sony.com}

\icmlkeywords{Music source separation, STFT phase}

\vskip 0.3in
]

\printAffiliationsAndNotice{}

\begin{abstract}
Music source separation with deep neural networks typically relies only on amplitude features. In this paper we show that additional phase features can improve the separation performance. Using the theoretical relationship between STFT phase and amplitude, we conjecture that derivatives of the phase are a good feature representation opposed to the raw phase. We verify this conjecture experimentally and propose a new DNN architecture which combines amplitude and phase. This joint approach achieves a better signal-to distortion ratio on the DSD100 dataset for all instruments compared to a network that uses only amplitude features. Especially, the bass instrument benefits from the phase information.
\end{abstract}

\section{Introduction}
\label{sec:intro}

\emph{Music source separation} (MSS) refers to the problem of obtaining instrument estimates $\mathbf{\hat s}_j(n) \in \R^I$ from the mixture
\begin{equation}
  \mathbf{x}(n) = \sum\nolimits_{j\in \mathcal{J}} \mathbf{s}_j(n),
\end{equation}
where $n$ denotes the discrete time index, $I$ gives the number of channels and $\mathcal{J}$ is the set of instruments. A common setup is the extraction of $\mathcal{J} := \{\text{bass}, \text{drums}, \text{vocals}, \text{other}\}$ from stereo mixtures, i.e., $I = 2$. This setup was used for the last \emph{SiSEC} contests on MSS \cite{ono20152015,liutkus20172016,stoter20182018} and is also the basis of our work.
 
This paper studies the appropriateness of the \emph{short-time Fourier transform} (STFT) phase as an input feature for MSS systems based on \emph{deep neural networks} (DNNs). Current state-of-the-art approaches perform MSS by only considering the mixture STFT amplitude from which they estimate the target instrument STFT amplitude \cite{huang2014deep,huang2014singing,uhlich2015deep,nugraha2016multichannel,uhlich2017improving,takahashi2017multi,takahashi2018mmdenselstm}. It is well known that the STFT phase contains useful information for speech enhancement, see e.g. \cite{gerkmann2015phase} and, therefore, should not be neglected. Recent attempts have been made in order to improve MSS using phase. \cite{lee2017fully} proposed a fully complex-valued DNN, which predicts the complex STFT of the target instrument from the complex mixture STFT. \cite{dubey2017does} analyzed whether phase is beneficial as input feature for a DNN compared to a network using only amplitude. Another approach is \cite{TAKIS2018}, which estimates the phase of the instrument from the mixture amplitude and phase by treating the phase retrieval problem as a classification problem. This paper presents another approach where phase is used as an additional input feature to improve the amplitude estimation. In contrast to \cite{dubey2017does}, we propose a special architecture to exploit the information that is present in the STFT phase as a simple concatenation of amplitude and phase at the input of the network yields trained networks that focus only on amplitude information.

\begin{figure*}[!ht]
  \centering
  \subfigure[Amplitude estimation from mixture amplitude.]{\resizebox{0.45\linewidth}{!}{\includegraphics{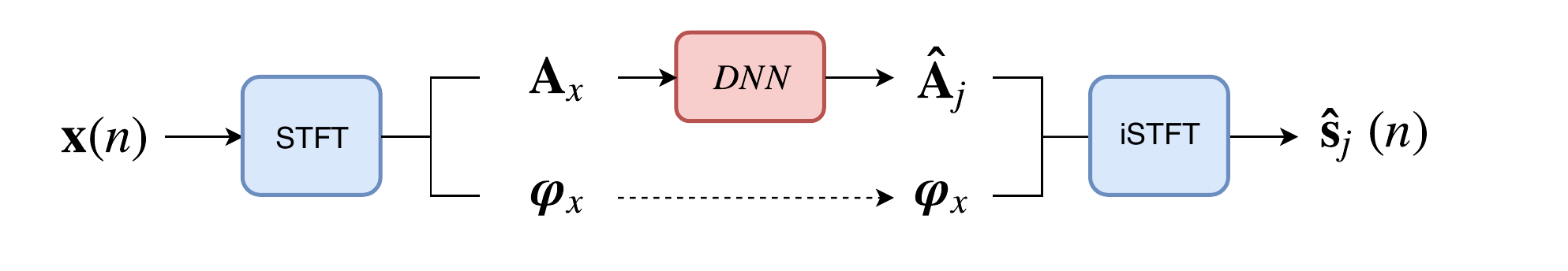}}}
  \hfill
  \subfigure[Amplitude estimation from mixture amplitude and phase.]{\resizebox{0.45\linewidth}{!}{\includegraphics{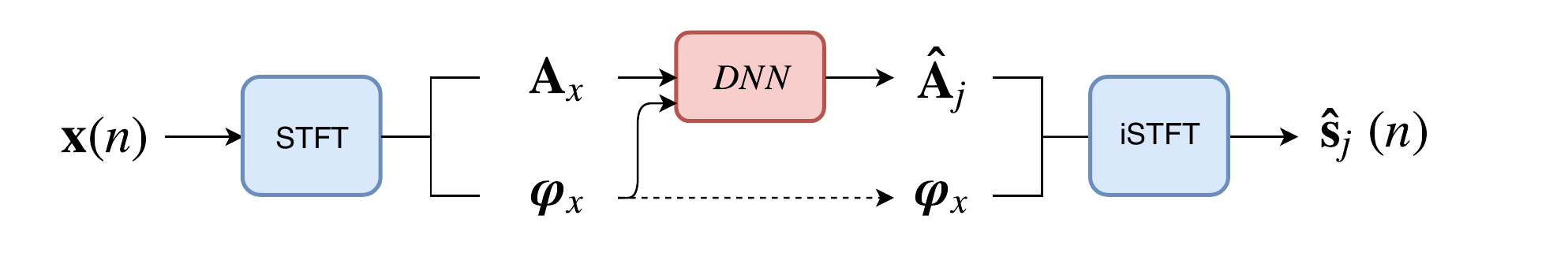}}}
  \subfigure[Amplitude estimation from mixture amplitude and phase estimation from mixture amplitude and phase.]{\includegraphics[width=0.45\linewidth]{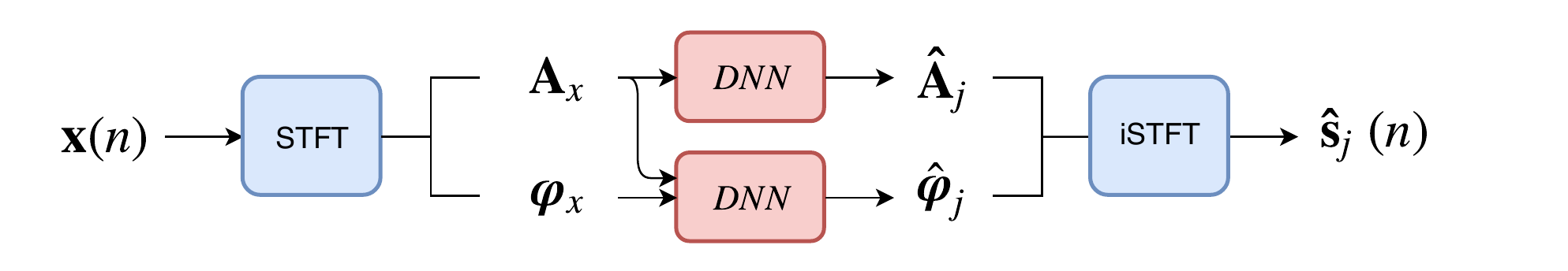}}
  \vskip -0.3cm
  \caption{Comparison of different MSS approaches.}
  \label{fig:comparison_mss_approaches}
  \vskip -0.3cm
\end{figure*}

We first show that expressing the phase through its \emph{instantaneous frequency} (time derivative) and its \textit{group delay} (frequency derivative) greatly improves the efficiency of DNNs compared to networks fed with raw phase inputs. This is done by looking at experimental results as well as studying the theoretical relationship between phase and amplitude of a continuous-time STFT. Moreover, we demonstrate that the discrete-time STFT introduces systematic shifts into the phase and that correcting these shifts improves the efficiency of the DNN to exploit these features.

Finally, we design a network architecture which takes full advantage of this additional feature. It is formed by two independent networks, taking respectively amplitude and phase, whose outputs are concatenated afterwards through a dense layer. Intuitively, each network independently extracts features from amplitude and phase and forwards them to a fusion layer, which reconstructs the spectrum based on these features.
With the suggested data pre-processing method and architecture design, our system achieves on average a relative improvement of 2.3\% and up to 6\% for bass compared to an amplitude-only system.

For clarity, the paper is divided into two parts. Sec.~\ref{sec:phase_feature} is dedicated to the properties of the phase as a feature for MSS. In this section, we consider the problem of using only the phase for estimating the instrument amplitude. This allows us to better understand this feature and the development of an appropriate pre-processing method. Sec.~\ref{sec:combination_of_amp_phase}, in contrast, considers both amplitude and phase from the mixture signal in order to produce an improved estimate of the instrument amplitude, which is the ultimate goal of the paper.

\begin{table}[t]
\caption{Comparison of upper limits reachable by two different approaches. Results on DSD100 test set (SDR in dB).}
\label{tab:approach_comparison}
\vskip 0.15in
\centering
\begin{small}
\begin{tabular}{lcccr}
\toprule
\textbf{\scriptsize Instrument} & \scriptsize \bf DNN$_\textbf{A}$ using  & \scriptsize \textbf{Upper baseline of} & \scriptsize  \textbf{Upper baseline of} \\
& \scriptsize \bf approach (a) & \scriptsize  \textbf{approach (b)} & \scriptsize  \textbf{approach (c)} \\
& & \tiny (\textbf{A$_\text{IRM}$ \& $\boldsymbol{\varphi}_\text{mixture}$}) & \tiny (\textbf{A$_\text{DNN}$ \& $\boldsymbol{\varphi}_\text{oracle}$}) \\
\midrule
Bass      & 3.24 & 7.92 		& 6.59 	\\
Drums   & 4.68 & 8.53 		& 6.15	\\
Other    & 3.54 & 8.19		& 5.35	\\
Vocals   & 4.78 & 11.10		& 6.86	\\
\bottomrule
\end{tabular}
\end{small}
\vskip -0.1in
\end{table}


\section{Phase as Input Feature}
\label{sec:phase_feature}

\subsection{Motivation}

Fig.~\ref{fig:comparison_mss_approaches} shows three different approaches for MSS, where $\mathbf{A}(k,m) \in \R_+^I$ and $\boldsymbol{\varphi}(k,m) \in [-\pi,\pi)^I$ denote the STFT amplitude and phase at frequency bin index $k$ and frame index $m$. $\mathbf{\hat s}_j(n) \in \R^I$ is the estimated target instrument signal.

Typically, approach (a) is used where the instrument amplitude $\mathbf{\hat{A}}_j$ is estimated from the mixture amplitude $\mathbf{A}_x$, while the instrument phase is simply approximated by the mixture phase $\boldsymbol{\varphi}_x$. The estimated instrument $\mathbf{\hat{s}}_j(n)$ is produced by applying an inverse STFT with the estimated source amplitude $\mathbf{\hat{A}}_j$ and mixture phase $\boldsymbol{\varphi}_x$.

Approaches (b) and (c) show two different ways to improve upon (a). Approach (b), which was, e.g., used in \cite{dubey2017does}, is similar in all respects except that the mixture phase is used to improve the instrument amplitude estimation. Approach (c), which was, e.g., followed by \cite{TAKIS2018}, estimates the instrument phase $\boldsymbol{\hat{\varphi}}_j$ which can then be used for the inverse STFT.

In order to choose between the two possible improvements, we did a simple experiment shown in Table~\ref{tab:approach_comparison}. We compare the upper limits achievable by both strategies: on one side a signal synthesized with the \emph{ideal ratio mask} (IRM) amplitude and the mixture phase; on the other side the oracle phase and the amplitude estimation from the network DNN$_\text{A}$\footnote{DNN$_\text{A}$ is a network which estimates the instrument amplitude from the mixture amplitude. Please refer to Sec.~\ref{sec:combination_of_amp_phase} for more details about this network.}. We can see that approach (b) has more room for improvement as the upper limit achievable has a relative improvement of 122\%. In contrast, the upper limit of approach (c) allows an average relative improvement of 57\%, which indicates that currently the amplitude estimation is still the main bottleneck for MSS performance. We therefore investigate approach (b) in this paper.

\subsection{Theoretical Relationship}
\label{subsec:theory}

Interestingly, for the continuous-time STFT
\begin{equation}
    X(\omega,t) = A(\omega,t)e^{j \varphi(\omega,t)}
\end{equation}
of a continuous-time signal $x(t)$, there is a theoretical relationship between the amplitude $A(\omega,t)$ and the phase $\varphi(\omega,t)$. The continuous-time STFT is given by
\begin{align}
  X(\omega,t) = e^{j\omega t/2}\int\limits_{-\infty}^\infty x(u)h(t-u)e^{-j\omega u} \text{d}u.
\end{align}

Using a Gaussian window $h(t) = \lambda^{-1/2}\pi^{-1/4}e^{-t^2/(2\lambda^2)}$, \cite{auger2012phase} showed that
\begin{subequations}
\begin{align}
  \frac{\partial}{\partial t}\varphi(\omega,t) &= \hphantom{-}\lambda^{-2}\frac{\partial}{\partial\omega} \log\left( A(\omega,t) \right) + \frac{\omega}{2},\\
	\frac{\partial}{\partial\omega}\varphi(\omega,t) &= -\lambda^{2}\frac{\partial}{\partial t} \log\left( A(\omega,t) \right) - \frac{t}{2}.
\end{align}\label{eq:relationship}\end{subequations}
From \eqref{eq:relationship}, we can see that the derivatives of phase and log-magnitude are linked and, therefore, we hope that the amplitude estimation for our target instrument from the mixture phase can be improved by using phase features.

Furthermore, we conjecture from \eqref{eq:relationship} that better results for the amplitude estimation can be obtained if we work with time/frequency derivatives of the phase instead of the raw phase. This intuition will be experimentally confirmed in Sec.~\ref{subsec:experiment}. As we work with discrete-time signals, we will approximate the derivatives by differences, i.e., in the following we will use
\begin{subequations}
\begin{align}
    \Delta_t\varphi &:= \varphi(k,m) - \varphi(k,m-1),\\
    \Delta_f\varphi &:= \varphi(k,m) - \varphi(k-1,m).
\end{align}
\end{subequations}
Please note that from the phase information, we are able to recover the amplitude up to an unknown scale. This can be seen from considering a signal $s(n) \in \C$ and a scaled version $s'(n) = a \cdot s(n)$ with $a > 0$ as $\angle s(n) = \angle s'(n)$ whereas $|s(n)| \neq |s'(n)|$. Hence, the phase only contains information about variations of the amplitude. This property is consistent with \eqref{eq:relationship} which links phase and log-amplitude through their derivatives.

\subsection{Shifts in discrete Short-Time Fourier Transform}
\label{subsec:shift}

\begin{figure}[!t]
  \centering
  {\includegraphics[width=\columnwidth, clip,trim=0 0 0 0]{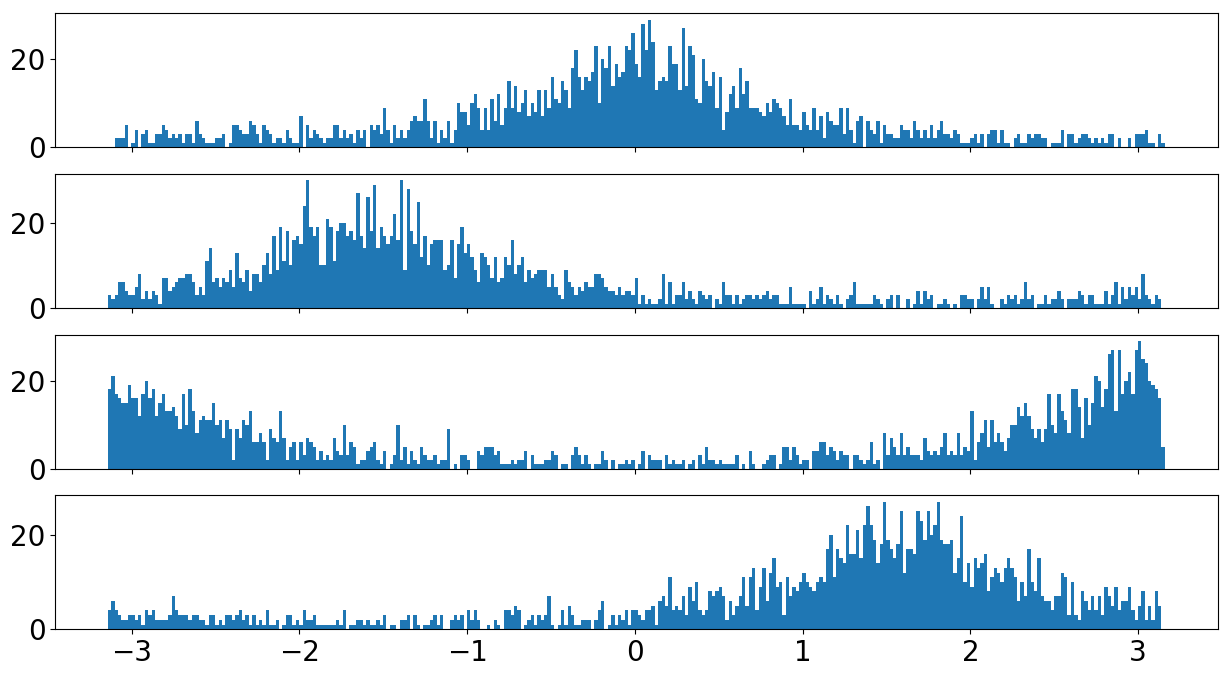}}
  	\vspace{-0.7cm}
	\caption{Distribution of \textit{instantaneous frequencies} ($\Delta_t \varphi$) over all time frames for one song from DSD100. From top to bottom, four successive frequency bins are considered and the histograms show a shift of $-k\frac{\pi}{2}$ introduced by the STFT with FFT size $N = window \ size = 4096$ and hop size $n_0 = 1024$ (overlap of $75\%$).}
	\label{fig:shift}
	\vspace{-0.3cm}
\end{figure}

Fig.~\ref{fig:shift} shows the distribution of $\Delta_t \varphi = \varphi(k,m) - \varphi(k,m-1)$ for consecutive frequency bins. We can observe a systematic offset in the statistical distribution which can be explained by the shift theorem of the \emph{discrete Fourier transform} (DFT) \cite{smith2007mathematics}. It states that a delay in the time domain results in a linear phase term in the frequency domain, i.e.,
\begin{align}
    x(n-n_0) \xrightarrow{DFT} e^{j\frac{2\pi}{N}k n_0} X(k),
\end{align}
where $n_0$ is the shift and $N$ the DFT/FFT size.

Therefore, in the case of a stationary signal transformed by an STFT, with hop size $n_0$, the phase of two consecutive frequency bins is expected to be shifted by a term
\begin{align}
\text{phase shift} = -\frac{2\pi}{N}k n_0 
\end{align}
For example, an overlap of $75\%$ results in a shift of $-k\frac{\pi}{2}$ which can also be seen in Fig.~\ref{fig:shift}.

DNNs are known to be sensitive to the feature distribution and, therefore, this shift should be properly compensated for during the pre-processing stage, as described in Sec.~\ref{subsec:preproc}, in order to ensure a proper training of the DNN.

\begin{figure*}[!ht]
  \hfil
  {\center
  \subfigure[Before shift correction ($\Delta_f \varphi$ and $\Delta_t \varphi$)]{\includegraphics[width=\columnwidth, clip,trim=0 0 0 0]{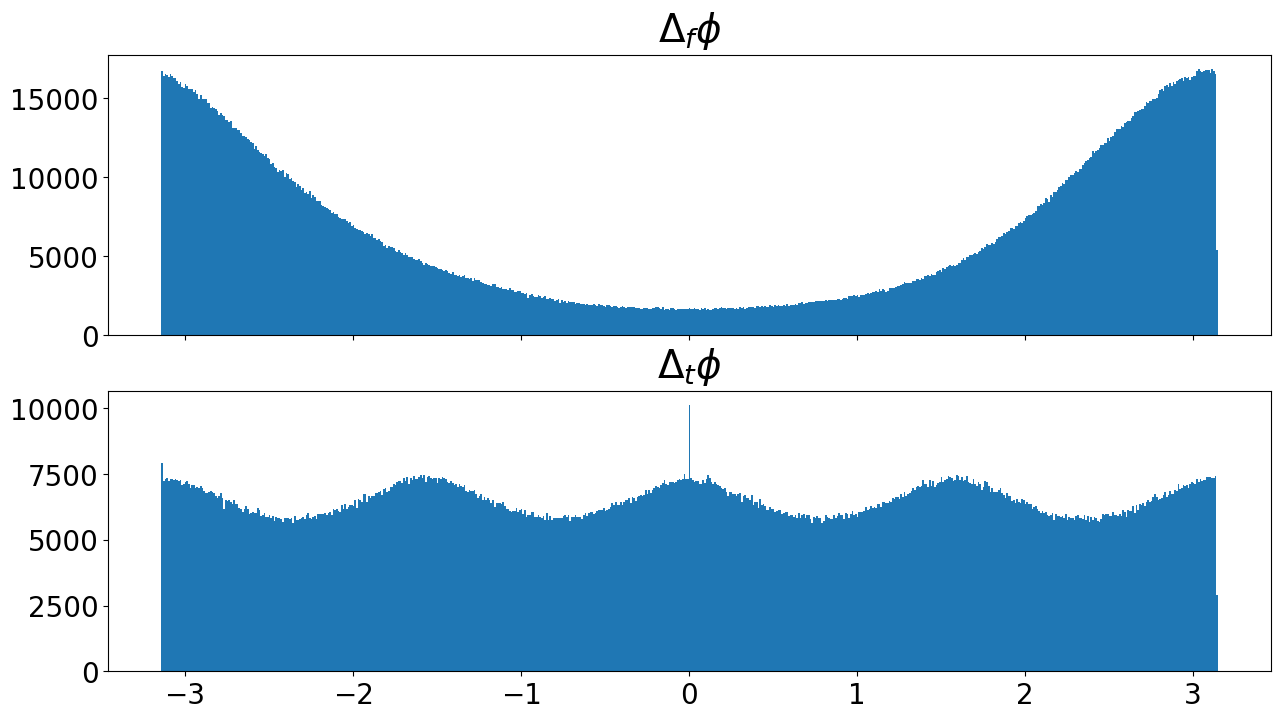}}
  \hfill
  \subfigure[After shift correction ($\Delta_f \varphi$ and $\Delta_t \varphi$)]{\includegraphics[width=\columnwidth, clip,trim=0 0 0 0]{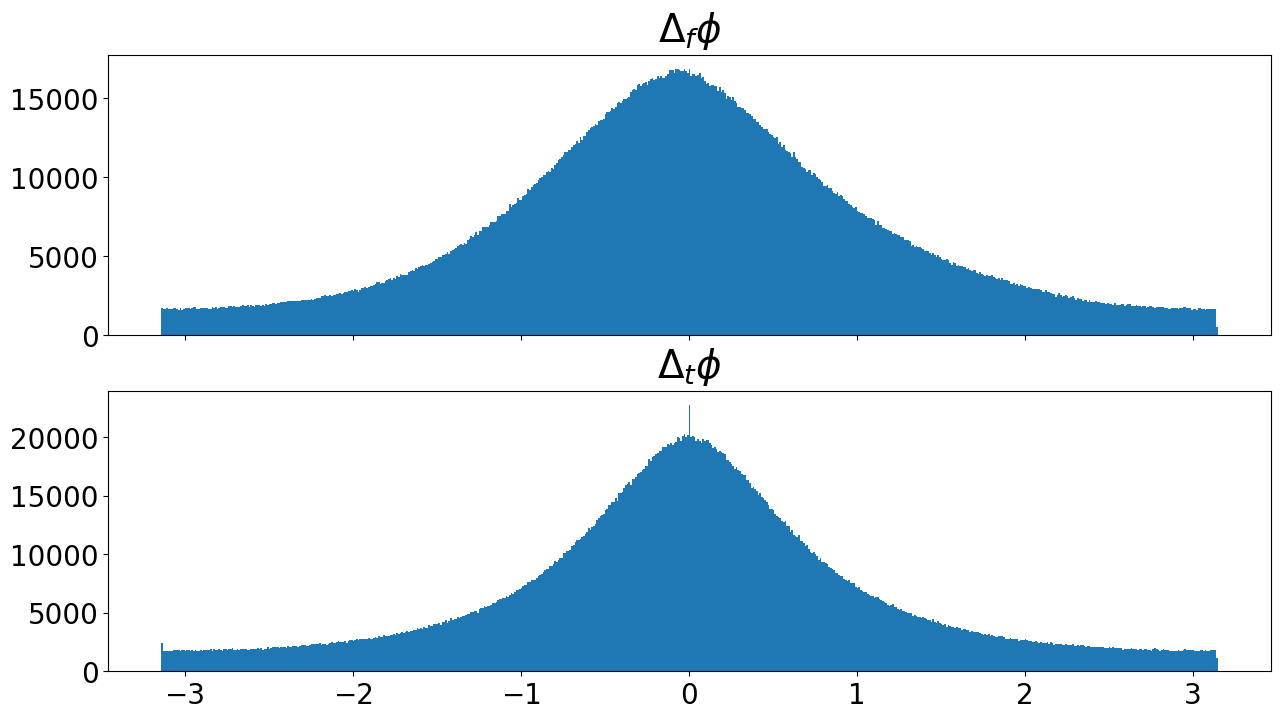}}
  }
  \caption{Statistical distribution of the group delays ($\Delta_f \varphi$) and instantaneous frequencies ($\Delta_t \varphi$) before and after shift correction.}
  \label{fig:distribution_phase}
\end{figure*}


\subsection{Pre-processing}
\label{subsec:preproc}

According to the conclusions drawn in Sec.~\ref{subsec:theory} and Sec.~\ref{subsec:shift}, we apply the following pre-processing steps to the raw phase:
\begin{itemize}
	\item The time and frequency derivatives are first approximated by the difference between two consecutive time frames ($\Delta_t \varphi$) and by the difference between two consecutive frequency bins ($\Delta_f \varphi$), respectively.
	\item A linear term $2 \pi k \frac{n_0}{N}$ is added to the time differences in order to compensate for the effect described in Sec.~\ref{subsec:shift}. Consequently, for a stationary signal $\Delta_t \varphi = 0$.
	\item For $\Delta_f \varphi$, we could empirically observe a systematic shift of $\pi$ in its statistical distribution, see Fig.~\ref{fig:distribution_phase}~(a). We compensate it by subtracting $\pi$ in order to obtain $\mathbb{E}(\Delta_f \varphi) = 0$.
	\item Finally, all values are wrapped to $[-\pi, \pi)$ using
    \begin{equation}
        \Delta \varphi = \left((\Delta \varphi + \pi) \bmod 2\pi \right)- \pi.
    \end{equation}
\end{itemize}
The effects of this pre-processing method on feature statistical distribution are illustrated in Fig.~\ref{fig:distribution_phase}.

\subsection{Experimental Validation}
\label{subsec:experiment}

In order to see whether our pre-processing is effective, we run two experiments, which we now describe in detail.

\begin{figure}[t]
  \centering
  \includegraphics[width=\columnwidth, clip,trim=50 0 10 0]{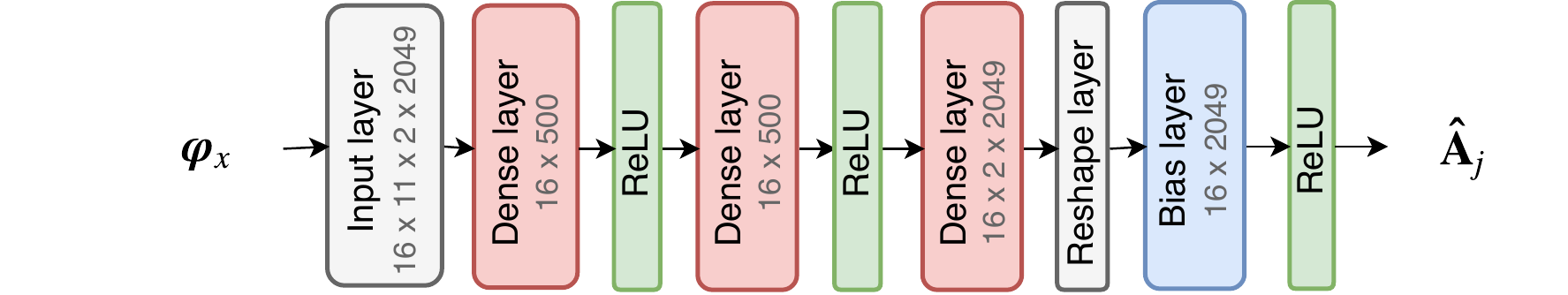}
  \vspace{-0.5cm}
  \caption{Network architecture with phase feature only.}    
  \label{fig:phase_only_arch}
  \vspace{-0.3cm}
\end{figure}

\begin{figure*}
  \centering
 {\includegraphics[width=0.8\linewidth]{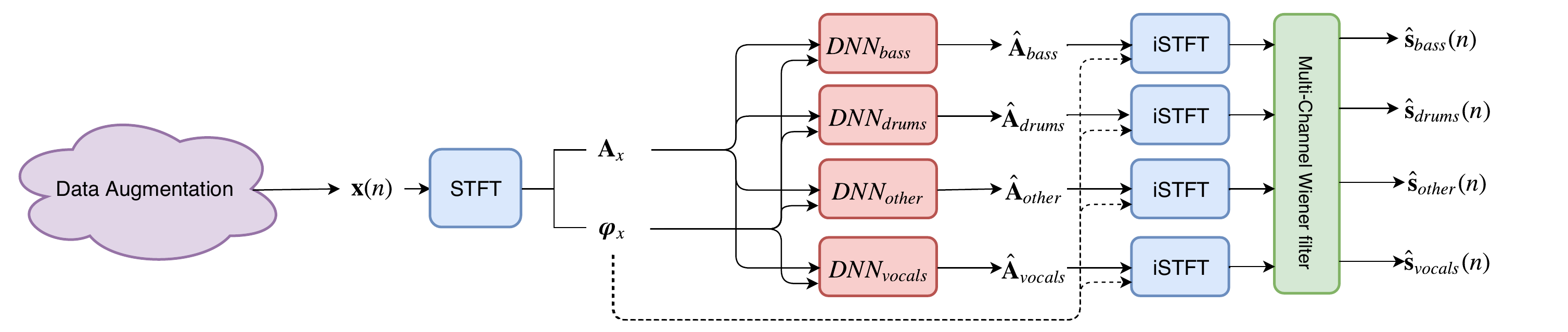}}
   \vspace{-0.1cm}
	\caption{Music source separation general framework. In Sec.~\ref{subsec:experiment}, the DNNs are only fed with the phase features in order to assess suggested pre-processing methods. Thus, with respect to the figure, the links between ${\bf A}_x$ and DNNs are removed. In Sec.~\ref{sec:combination_of_amp_phase}, both amplitude and phase features are input to the DNNs as depicted in the figure.}
	\label{fig:framework}
\end{figure*}

The network used to evaluate the suggested pre-processing method is formed by two dense layers of 500 hidden units, intersected by ReLU non-linearities and completed by a dense output layer matching the target dimensions. At the very end, a bias layer initialized with the average amplitude per frequency bin over the training set shifts the output and a ReLU non-linearity ensures non-negative output values. We use a context of five preceding/succeeding frames as temporal context. Fig.~\ref{fig:phase_only_arch} shows the network structure and the overall MSS framework is summarized in Fig.~\ref{fig:framework}.

We now give the results for estimating the STFT amplitude from the phase using a DNN. By these experiments, we show that it is advantageous to consider the time/frequency derivatives instead of the raw phase. Furthermore, the pre-processing described in Sec.~\ref{subsec:preproc}, is also shown to be relevant.

\begin{figure*}
  \centering
  {\includegraphics[width=0.98\linewidth]{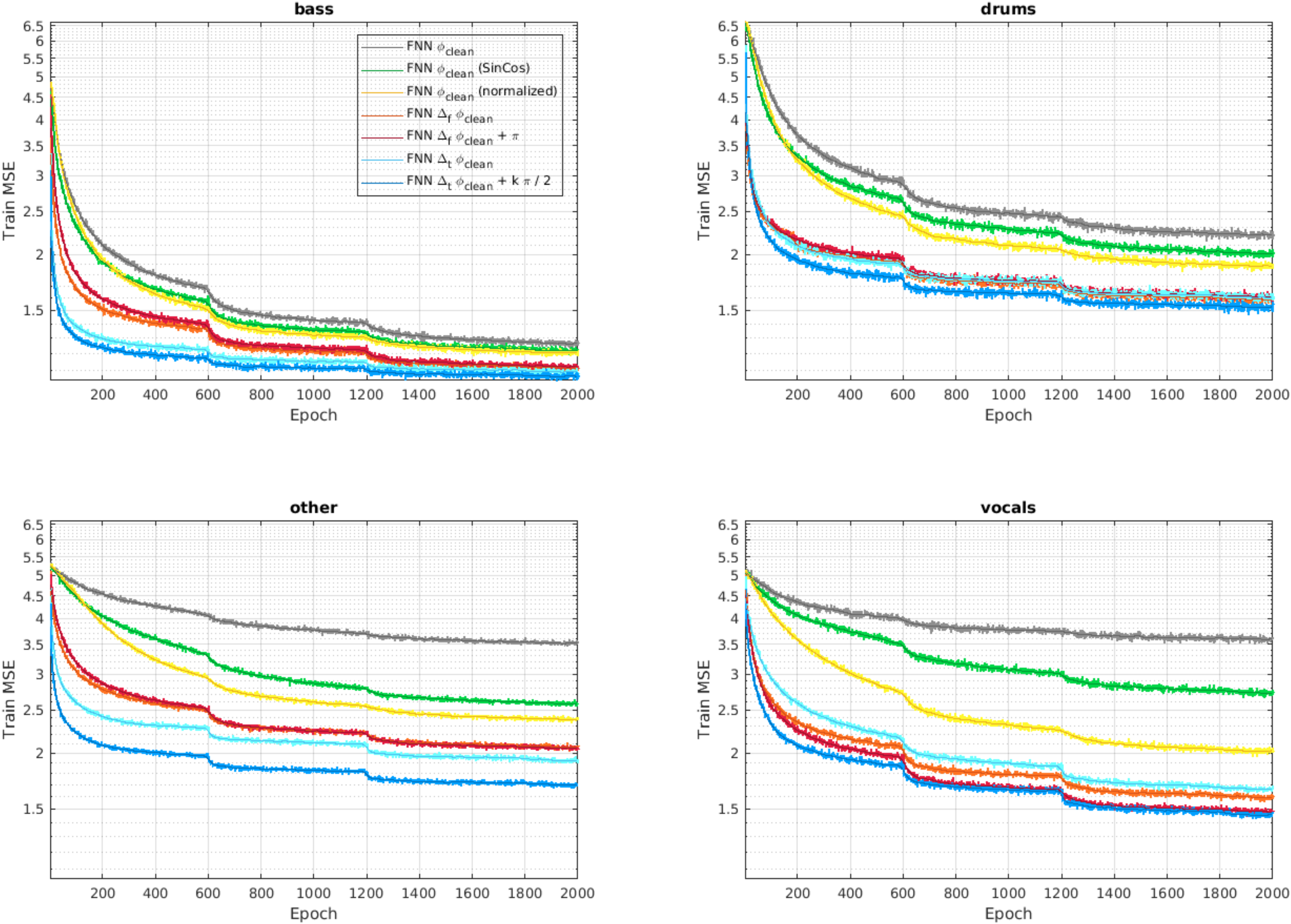}}
   \vspace{-0.1cm}
	\caption{Training MSE for reconstructing amplitude from phase. Suggested pre-processing methods drastically improve the learning power of the network in comparison with raw phase. In particular, instantaneous frequencies and group delays are good representations. Moreover, shift correction systematically improves performances for all instrument.}
	\label{fig:train_MSE_clean}
	\vskip -0.2cm
\end{figure*}

In the first experiment, we reconstruct the instrument amplitude from the instrument phase. Thus, we do not consider a separation problem but instead focus on the ability of a DNN to recover a signal knowing its phase. By this, we can compare different phase feature representations and observe their effects on the DNN learning power. The training MSE curves are shown in Fig.~\ref{fig:train_MSE_clean}. We can observe that the phase derivatives show the best performance as we previously conjectured.

\begin{figure}
  \centering
  {\includegraphics[width=\columnwidth]{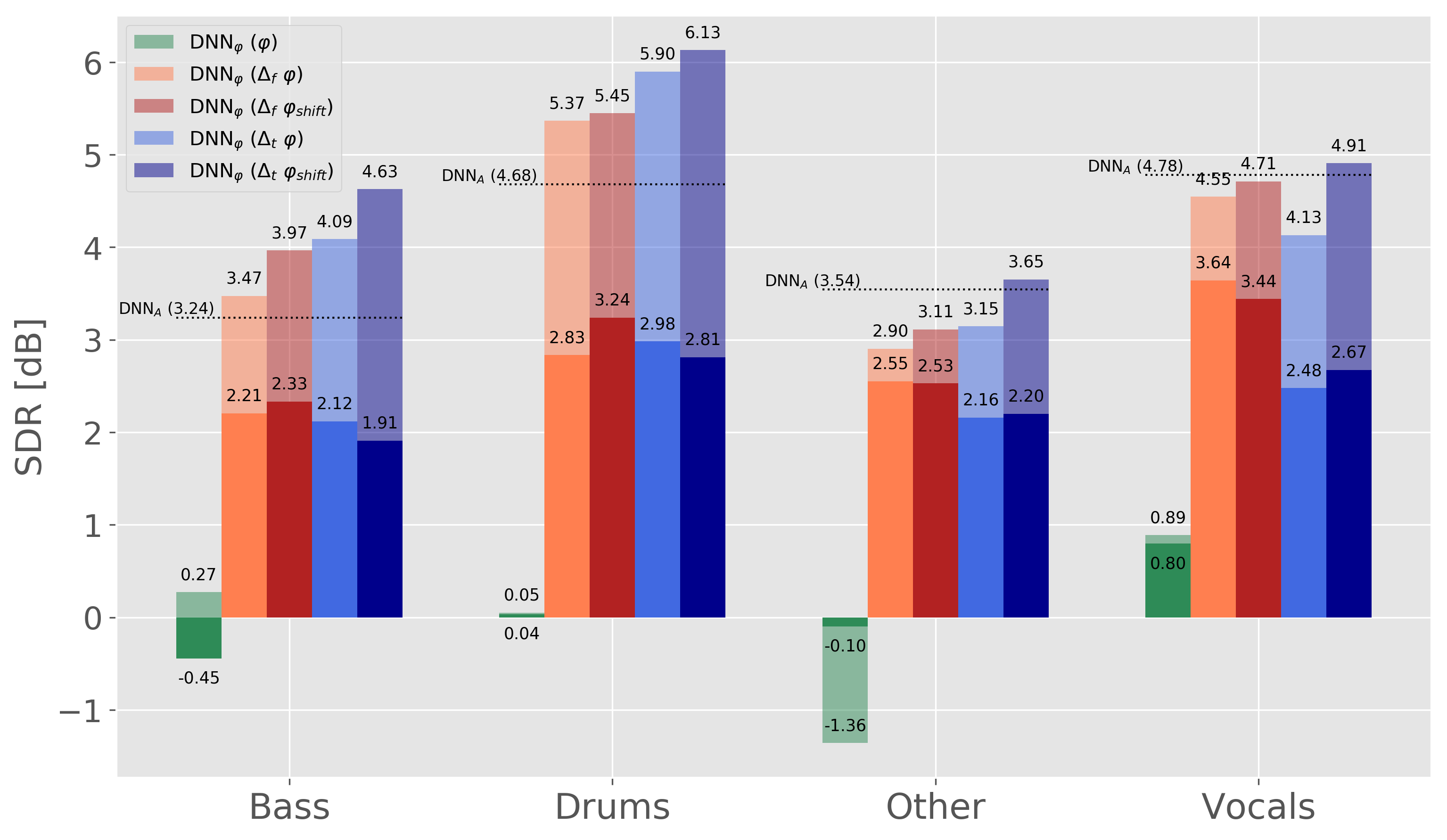}}
  \vspace{-0.5cm}
  \caption{SDR results for phase only networks on DSD100. Comparison between various pre-processing methods. Note the poor results of raw phase without pre-processing on green. Each bar represents the accuracy obtained with a certain pre-processing method for a certain instrument. Height of opaque bar states the score on test set, while height of transparent bar (typically higher) is the score on training set. Horizontal dotted lines over each instrument give the score obtained by a DNN$_\text{A}$ on the test set.}
	\label{fig:SDR_phase_only}
	\vskip -0.2cm
\end{figure}

In the second experiment, we estimate the instrument amplitude from the mixture phase. This goes one step further than the previous experiment by involving separation in the comparative analysis of the pre-processing methods. The trained networks are then integrated in the MSS framework illustrated in Fig.~\ref{fig:framework}. Estimations are scored following SiSEC 2016 policy \cite{liutkus20172016}. Fig.~\ref{fig:SDR_phase_only} shows the \emph{signal-to-distortion ratio} (SDR) values \cite{vincent2007first} on the DSD100 dataset where the values are obtained by first averaging the SDR values for each song and then computing the median over all $50$ songs of the train set or test set, respectively. Again, we can observe that phase derivatives are a much better feature representation and that shift correction systematically improves learning power of the system, leading occasionally to overfitting. The best test SDR is achieved by the frequency-derivative representation of the phase which generalizes better than the time-derivative representation.

Note that a network fed with phase features can only estimate the amplitude values up to a scale, meaning that it uses the average amplitude value per frequency bin of the training set as a starting point and estimates the variations from it based on the phase input. The post-processing stage uses a multi-channel Wiener filter \cite{sivasankaran2015robust,nugraha2016multichannel,uhlich2017improving} to recover the correct scale afterwards.

\section{Combining Amplitude and Phase Features}
\label{sec:combination_of_amp_phase}

In the previous section, we have seen that phase features can be used to estimate the instrument STFT amplitude. Therefore, we now turn to the problem of combining phase and amplitude features.

\subsection{Proposed Architecture}
\label{subsec:combine:arch}

The most straight-forward way of combining amplitude and phase is a concatenation of both features at the input of the DNN. However, training such an approach results in networks that only rely on amplitude features as they set all weights in the input layer corresponding to the phase close to zero.\footnote{In our opinion, this behaviour is due to the fact that the information in the STFT mixture amplitude is more easily accessible than the information in the STFT phase.} We could observe this if we use the raw phase as well as if we use the phase pre-processing described in Sec. \ref{subsec:preproc}. 

\begin{figure}
  \vskip 0.2in
  \centering
  {\includegraphics[width=\columnwidth, clip,trim=60 0 20 0]{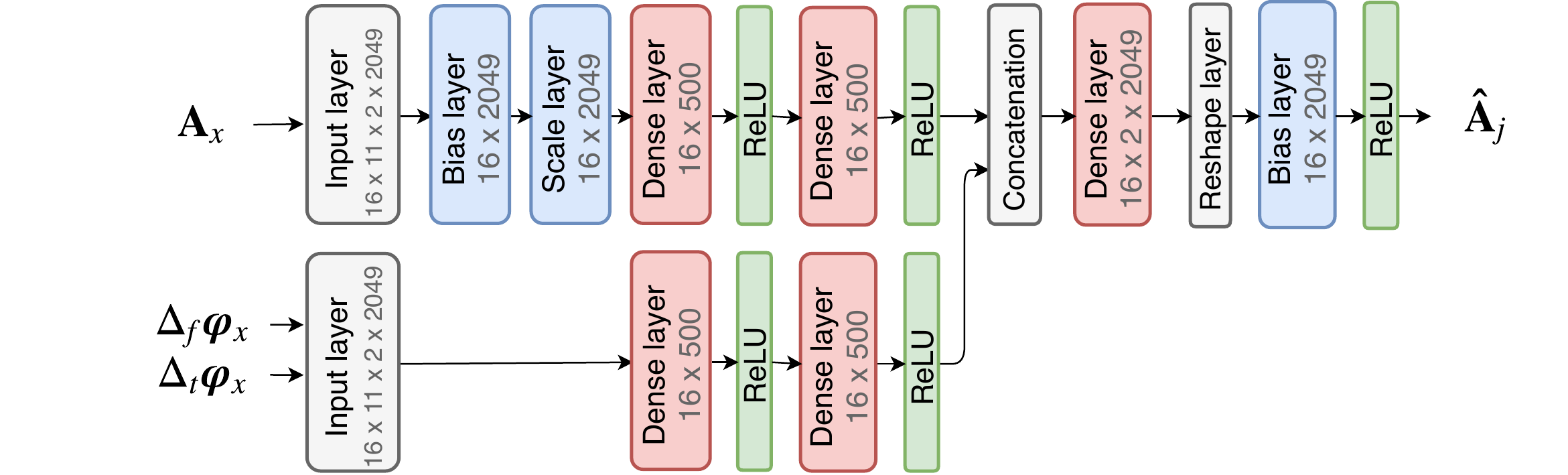}}
  \vskip -0.2cm
	\caption{Proposed architecture for combining amplitude and phase.}
	\label{fig:concat_architecture}
  \vskip -0.3cm
\end{figure}

Hence, we have to take special care to exploit the information of the phase features and we use the DNN architecture that is shown in Fig.~\ref{fig:concat_architecture}. Instead of concatenating the features directly, we first process both through two dense layers before concatenating them.

The upper part of the network in Fig.~\ref{fig:concat_architecture} deals with the amplitude features. The features are first normalized by a bias layer and a scale layer, initialized with the mean and standard deviation per frequency bin over the training set. Two fully connected layers of 500 hidden units perform the feature extraction. The lower part of the network in Fig.~\ref{fig:concat_architecture} deals with the phase features. It takes as input both time and frequency derivatives, properly pre-processed as described in Sec.~\ref{subsec:preproc} and stacked together into an extra dimension. As for amplitude, two fully connected layers of 500 hidden units perform the feature extraction. The concatenation layer stacks the output of both previous networks and produces the amplitude estimates, which are de-normalized with the help of a final bias layer and a ReLU non-linearity, as described in Sec.~\ref{subsec:experiment}. The training process is similar to the one described in Sec.~\ref{subsec:experiment}. Time context is, as well, kept to five preceding/succeeding frames.

\subsection{Results}

\begin{figure}
  \centering
  \resizebox{\linewidth}{!}{\includegraphics{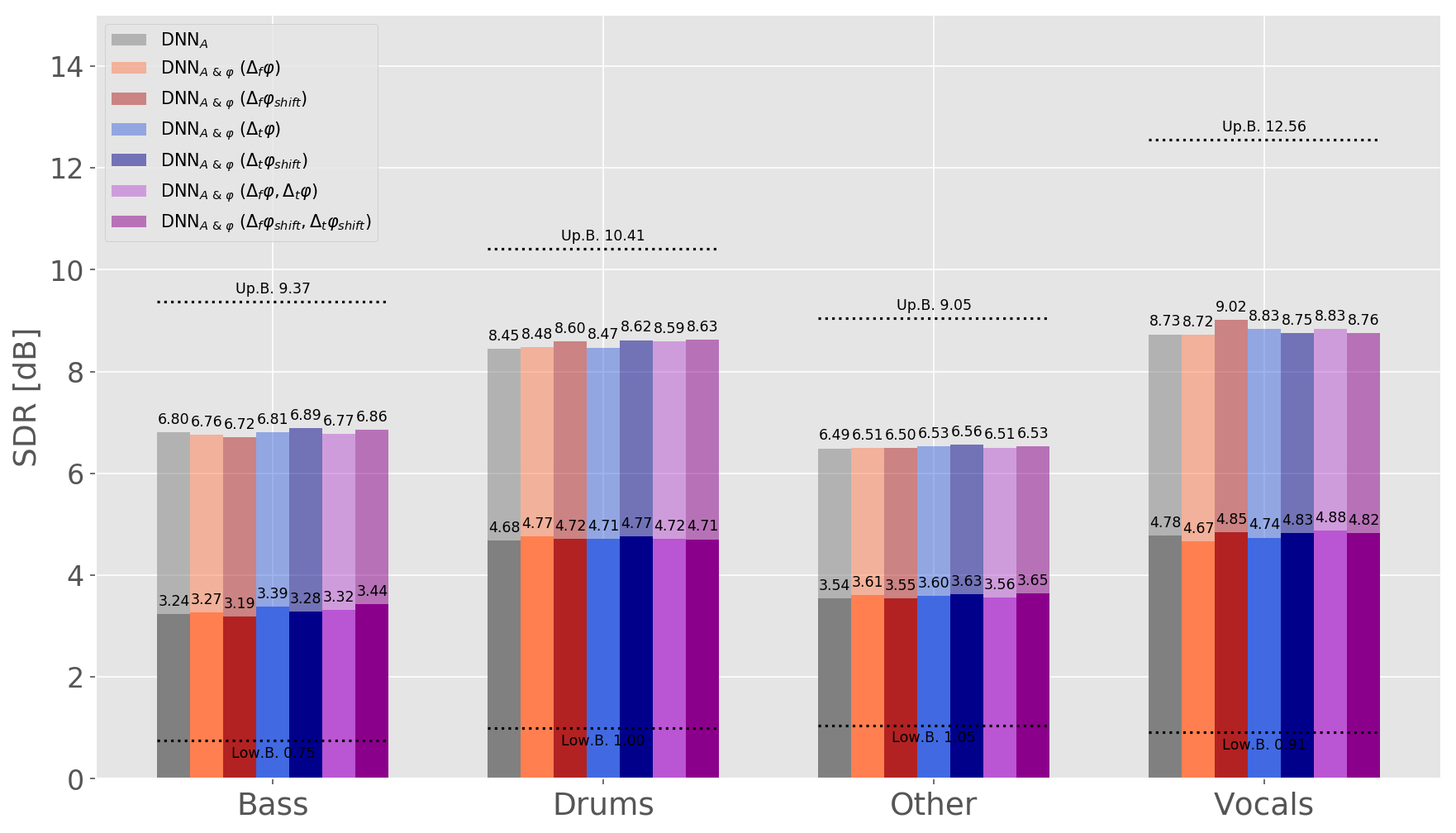}}
	\vspace{-0.5cm}
	\caption{SDR results for networks combining phase and amplitude on DSD100. Comparison between various pre-processing methods. Each bar represents the accuracy obtained with a certain pre-processing method for a certain instrument. Height of opaque bar states the score on test set, while height of transparent bar (typically higher) is the score on training set. First score of each instrument, in gray, is given for illustrative purpose only and show the score obtained by DNN$_\text{A}$.}
	\label{fig:SDR_amp_phase}
	\vspace{-0.3cm}
\end{figure}

Fig.~\ref{fig:SDR_amp_phase} shows the results obtained with amplitude and phase combination. For comparison, we also trained a network DNN$_\text{A}$, which uses only amplitude as feature and has the same structure as shown in Fig.~\ref{fig:concat_architecture} with the phase branch removed. Therefore, the amplitude information undergoes the same transformations and we can directly compare the two networks.

We use different combinations of pre-processing methods described in \ref{subsec:preproc} in order to experience the individual relevance of each step. As expected, applying all proposed phase pre-processing methods together is beneficial for the MSS performance.

Finally, Table~\ref{tab:concat_results} shows the SDR obtained on the DSD100 test set. Comparing the baseline system DNN$_\text{A}$ with DNN$_\text{A \& $\varphi$} \ (\Delta_f \varphi_\text{shift}, \Delta_t \varphi_\text{shift})$, we observe that we can improve the SDR for all instruments and that especially the bass instrument improves by $0.2$ dB.

\begin{table}[t]
\caption{Results for concatenation architecture on DSD100 test set (SDR in dB).}
\label{tab:concat_results}
\vskip 0.15in
\centering
\begin{small}
\begin{tabular}{lcccr}
\toprule
\textbf{Instrument} & \textbf{DNN$_\text{A}$} & \textbf{DNN$_\text{A \& $\boldsymbol\varphi$}$} & \textbf{Relative improv.} \\
\midrule
Bass      & 3.24 		& 3.44 		& $+ 6.17 \%$ \\
Drums     & 4.68 		& 4.71		& $+ 0.64 \%$ \\
Other     & 3.54		& 3.65		& $+ 3.11 \%$ \\
Vocals    & 4.78		& 4.82		& $+ 0.84 \%$ \\
\bottomrule
\end{tabular}
\end{small}
\end{table}

\section{Conclusion}

In this paper, we proposed to consider the phase as an additional input feature to enhance the amplitude estimation. We studied the relationship between the phase and the amplitude of an STFT and deducted a meaningful pre-processing, which was experimentally confirmed as relevant. We also found that special care must be taken in order to combine phase and amplitude features and, consequently, designed an adequate network architecture. The developed system improved SDRs on DSD100 for all instruments compared to an amplitude-only network with a similar network structure which showed the effectiveness of our system. Perceptually, this results in instruments more clearly separated from each other.

\bibliography{references}

\begin{thebibliography}{18}
\providecommand{\natexlab}[1]{#1}
\providecommand{\url}[1]{\texttt{#1}}
\expandafter\ifx\csname urlstyle\endcsname\relax
  \providecommand{\doi}[1]{doi: #1}\else
  \providecommand{\doi}{doi: \begingroup \urlstyle{rm}\Url}\fi

\bibitem[Auger et~al.(2012)Auger, Chassande-Mottin, and
  Flandrin]{auger2012phase}
Auger, F., Chassande-Mottin, {\'E}., and Flandrin, P.
\newblock On phase-magnitude relationships in the short-time {Fourier}
  transform.
\newblock \emph{IEEE Signal Processing Letters}, 19\penalty0 (5):\penalty0
  267--270, 2012.

\bibitem[Dubey et~al.(2017)Dubey, Kenyon, Carlson, and Thresher]{dubey2017does}
Dubey, Mohit, Kenyon, Garrett, Carlson, Nils, and Thresher, Austin.
\newblock Does phase matter for monaural source separation?
\newblock \emph{arXiv preprint arXiv:1711.00913}, 2017.

\bibitem[Gerkmann et~al.(2015)Gerkmann, Krawczyk-Becker, and
  Le~Roux]{gerkmann2015phase}
Gerkmann, Timo, Krawczyk-Becker, Martin, and Le~Roux, Jonathan.
\newblock Phase processing for single-channel speech enhancement: History and
  recent advances.
\newblock \emph{IEEE Signal Processing Magazine}, 32\penalty0 (2):\penalty0
  55--66, 2015.

\bibitem[Huang et~al.(2014{\natexlab{a}})Huang, Kim, Hasegawa-Johnson, and
  Smaragdis]{huang2014deep}
Huang, Po-Sen, Kim, Minje, Hasegawa-Johnson, Mark, and Smaragdis, Paris.
\newblock Deep learning for monaural speech separation.
\newblock \emph{Proc. ICASSP}, pp.\  1562--1566, 2014{\natexlab{a}}.

\bibitem[Huang et~al.(2014{\natexlab{b}})Huang, Kim, Hasegawa-Johnson, and
  Smaragdis]{huang2014singing}
Huang, Po-Sen, Kim, Minje, Hasegawa-Johnson, Mark, and Smaragdis, Paris.
\newblock Singing-voice separation from monaural recordings using deep
  recurrent neural networks.
\newblock \emph{International Society for Music Information Retrieval
  Conference (ISMIR)}, 2014{\natexlab{b}}.

\bibitem[Lee et~al.(2017)Lee, Wang, Wang, Wang, and Wu]{lee2017fully}
Lee, Yuan-Shan, Wang, Chien-Yao, Wang, Shu-Fan, Wang, Jia-Ching, and Wu,
  Chung-Hsien.
\newblock Fully complex deep neural network for phase-incorporating monaural
  source separation.
\newblock In \emph{Acoustics, Speech and Signal Processing (ICASSP), 2017 IEEE
  International Conference on}, pp.\  281--285. IEEE, 2017.

\bibitem[Liutkus et~al.(2017)Liutkus, St{\"o}ter, Rafii, Kitamura, Rivet, Ito,
  Ono, and Fontecave]{liutkus20172016}
Liutkus, A., St{\"o}ter, F.-R., Rafii, Z., Kitamura, D., Rivet, B., Ito, N.,
  Ono, N., and Fontecave, J.
\newblock The 2016 signal separation evaluation campaign.
\newblock In \emph{Proc. LVA/ICA}, pp.\  323--332. Springer, 2017.

\bibitem[Nugraha et~al.(2016)Nugraha, Liutkus, and
  Vincent]{nugraha2016multichannel}
Nugraha, A.~A., Liutkus, A., and Vincent, E.
\newblock Multichannel music separation with deep neural networks.
\newblock In \emph{Proc. EUSIPCO}, pp.\  1748--1752, 2016.

\bibitem[Ono et~al.(2015)Ono, Rafii, Kitamura, Ito, and Liutkus]{ono20152015}
Ono, Nobutaka, Rafii, Zafar, Kitamura, Daichi, Ito, Nobutaka, and Liutkus,
  Antoine.
\newblock The 2015 signal separation evaluation campaign.
\newblock In \emph{International Conference on Latent Variable Analysis and
  Signal Separation}, pp.\  387--395. Springer, 2015.

\bibitem[Sivasankaran et~al.(2015)Sivasankaran, Nugraha, Vincent,
  Morales-Cordovilla, Dalmia, Illina, and Liutkus]{sivasankaran2015robust}
Sivasankaran, Sunit, Nugraha, Aditya~Arie, Vincent, Emmanuel,
  Morales-Cordovilla, Juan~A, Dalmia, Siddharth, Illina, Irina, and Liutkus,
  Antoine.
\newblock Robust {ASR} using neural network based speech enhancement and
  feature simulation.
\newblock In \emph{IEEE Workshop on Automatic Speech Recognition and
  Understanding (ASRU)}, pp.\  482--489. IEEE, 2015.

\bibitem[Smith(2007)]{smith2007mathematics}
Smith, Julius~O.
\newblock \emph{Mathematics of the Discrete Fourier Transform (DFT)}.
\newblock W3K Publishing, 2007.
\newblock ISBN 978-0-9745607-4-8.

\bibitem[St{\"o}ter et~al.(2018)St{\"o}ter, Liutkus, and Ito]{stoter20182018}
St{\"o}ter, F.-R., Liutkus, A., and Ito, N.
\newblock The 2018 signal separation evaluation campaign.
\newblock \emph{arXiv preprint arXiv:1804.06267}, 2018.

\bibitem[Takahashi \& Mitsufuji(2017)Takahashi and
  Mitsufuji]{takahashi2017multi}
Takahashi, N. and Mitsufuji, Y.
\newblock Multi-scale multi-band {DenseNets} for audio source separation.
\newblock In \emph{Proc. WASPAA}, pp.\  21--25. IEEE, 2017.

\bibitem[Takahashi et~al.(2018{\natexlab{a}})Takahashi, Agrawal, Goswami, and
  Mitsufuji]{TAKIS2018}
Takahashi, N., Agrawal, P., Goswami, N., and Mitsufuji, Y.
\newblock Phasenet: Discretized phase modeling with deep neural networks for
  audio source separation.
\newblock In \emph{Proc. Interspeech}, 2018{\natexlab{a}}.

\bibitem[Takahashi et~al.(2018{\natexlab{b}})Takahashi, Goswami, and
  Mitsufuji]{takahashi2018mmdenselstm}
Takahashi, N., Goswami, N., and Mitsufuji, Y.
\newblock {MMDenseLSTM}: {A}n efficient combination of convolutional and
  recurrent neural networks for audio source separation.
\newblock \emph{accepted for IWAENC}, 2018{\natexlab{b}}.

\bibitem[Uhlich et~al.(2015)Uhlich, Giron, and Mitsufuji]{uhlich2015deep}
Uhlich, S., Giron, F., and Mitsufuji, Y.
\newblock Deep neural network based instrument extraction from music.
\newblock In \emph{Proc. ICASSP}, pp.\  2135--2139. IEEE, 2015.

\bibitem[Uhlich et~al.(2017)Uhlich, Porcu, Giron, Enenkl, Kemp, Takahashi, and
  Mitsufuji]{uhlich2017improving}
Uhlich, S., Porcu, M., Giron, F., Enenkl, M., Kemp, T., Takahashi, N., and
  Mitsufuji, Y.
\newblock Improving music source separation based on deep neural networks
  through data augmentation and network blending.
\newblock In \emph{Proc. ICASSP}, pp.\  261--265. IEEE, 2017.

\bibitem[Vincent et~al.(2007)Vincent, Sawada, Bofill, Makino, and
  Rosca]{vincent2007first}
Vincent, Emmanuel, Sawada, Hiroshi, Bofill, Pau, Makino, Shoji, and Rosca,
  Justinian~P.
\newblock First stereo audio source separation evaluation campaign: data,
  algorithms and results.
\newblock In \emph{International Conference on Independent Component Analysis
  and Signal Separation}, pp.\  552--559. Springer, 2007.

\end{thebibliography}
\bibliographystyle{icml2018}

\end{document}